\documentclass[prd,aps,twocolumn,nofootinbib,preprintnumbers]{revtex4-1}
\usepackage{amsmath,amssymb,epsf}
\usepackage{graphicx}
\usepackage{natbib}
\newcommand{\etal}{it et. al.}

\begin{document}

%\begin{flushleft}
%%\hfill HRI/ST1402
%\hfill SINP/TNP/2014/19 
%\end{flushleft}
%\hfill{HRI/ST1402} 
%\hfill{ SINP/TNP/2014/XX}

%\title{Helping hands to pNGB dark energy}
\title{Viable dark energy models using pseudo-Nambu-Goldstone bosons}

\author{Debabrata Adak}
\email{debabrata.adak@saha.ac.in}
\affiliation{Astroparticle Physics and Cosmology Division\\
Saha Institute of Nuclear Physics, 1/AF Bidhannagar Kolkata-700064, India}

\author{Koushik Dutta}
\email{koushik.dutta@saha.ac.in}
\affiliation{Theory Division\\ Saha Institute of Nuclear Physics,
1/AF Bidhannagar Kolkata-700064, India}

%\date{\today}

\begin{abstract}
A canonical pseudo-Nambu Goldstone Boson (pNGB) can play the role of a 
dark energy field responsible for present cosmic acceleration. Confronting 
with the recent cosmological data, we find that the pNGB field requires 
spontaneous symmetry breaking scale $f$ close to $M_P$ and the initial 
field value fine-tuned. It is difficult to achieve a large $f$ in a 
theoretically consistent set-up. A possible resolution can be achieved 
by increasing the Hubble friction in well motivated particle physics models 
in the general set-up of modified gravity theories. 
We show two phenomenological examples of this set-up where the 
standard pNGB action have been modified by introducing terms motivated 
from galileon cosmology. We confront those examples with the recent 
supernovae, PLANCK and BAO data. We find that moderate values 
of the dimensionless constants that 
increase the friction, make $f << M_{P}$ and the generic initial conditions 
also favourable by the data.
We also comment how the fifth force constraints arising in these 
modified theories can be evaded.
\end{abstract}
\maketitle

\section{Introduction}
\label{intro}
Presently we live in a strange Universe where the recent observations suggest that the Universe has started to accelerate very recently \cite {Spergel:2006hy, Riess:1998cb, Perlmutter:1998np}. With the assumption of the General Theory of Relativity being the correct description of physical phenomenon at the cosmic scale, and the effects of inhomogeneities can be neglected, recent cosmic acceleration can be explained by the addition of some exotic matter with negative pressure (broadly called ``dark energy'') that has started to dominate the total energy budget of the Universe. The cosmological constant ($\Lambda$) is the simplest source of dark energy that can solve this problem. Moreover, the single parameter solution of introducing $\Lambda$ in the Einstein-Hilbert action is an excellent fit to the all observational data \cite{Ade:2013zuv}. But the required value of $\Lambda$ to match the observations is very tiny: $\Lambda\approx(2\times10^{-3} ~eV)^4$. Even if the cosmological constant is the reason behind the cosmic acceleration, its tiny value is extremely difficult to justify from the present theoretical understanding of the particle physics. Additionally, in the context of cosmological constant, there is no dynamical resolution to the question of why the Universe is getting dominated by the cosmological constant very recently.  Surely, the anthropic arguments in the context of String Theory landscape scenario is a possibility \cite{Weinberg:1987dv}.

Another well motivated idea of the source of dark energy is the existence 
of a scalar field $\phi$ (`quintessence') whose present energy density 
is approximately the above mentioned value of $\Lambda$ 
\cite{Wetterich:1987fm, Ratra:1987rm, Caldwell:1997ii}. For reviews on 
dark energy see \cite{Copeland:2006wr}, and for its effective field theory
approach see \cite{Bloomfield:2011np,2013PhRvD..87h3504M}.
The scalar field $\phi$ being 
dynamical, in contrast to the cosmological constant, its pressure $p$ is 
related to the energy density $\rho$ by an equation of state parameter 
$w \neq -1$ where $p = \omega \rho$. The fact that the cosmological constant 
(with $w = -1$) is still a very good fit to the data forces $w$ for the 
scalar field not being too away from $-1$. This essentially translates 
to a very flat potential for $\phi$ with its mass around $10^{-33}$ eV. 

Even though the idea of quintessence is very well motivated, finding a natural candidate for $\phi$ in particle physics is very challenging. It is due to the following two reasons. Firstly, quantum corrections due to the other fields in any theory will generically spoil the flatness of the potential \cite{Kolda:1998wq}. Secondly, the ultra-light $\phi$ field would carry fifth-force, typically of gravitational strength \cite{Carroll:1998zi, Chiba:1999wt}. All the models of dark energy are plagued with these problems, unless a symmetry protects its mass and the fifth-force constraints are avoided by some mechanisms.
%e.g. chameleon \ref{}, symmetron \ref{} or Vainstein \ref{}. 

These problems can be easily solved for the case of a pNGB potential whose mass is protected by a shift symmetry \cite{Frieman:1995pm}. The breaking of the symmetry in a controlled way allows us to keep the mass of $\phi$ radiatively stable and ``naturally'' light. A pNGB field can only couple to other fields with its derivative couplings, and it naturally suppresses the fifth-force constraints. A nice feature of the pNGB potential is that it has only two free parameters: parameter $f$ is related to the shift symmetry breaking scale, and $\mu$ is related to the explicit symmetry breaking scale. The present value of the cosmological constant fixes the value of $\mu$. Because of the periodic nature of the pNGB potential, the initial field range is compact ranging from $0$ to $2\pi$. In addition, $f$ can not be larger than the reduced Planck mass $M_P$, allowing us to constrain or possibly rule out the potential completely by confronting with the data. Following this line of thought, a detailed study of the parameter space of single pNGB potential was done in \cite{Dutta:2006cf}. It was found that unless the initial field values are chosen in a fine-tuned way, the data always prefer $f$ to be being close to $M_P$. In this work, we will reanalyse the situation in the light of the latest available cosmological data and reassure ourselves that the constraints have indeed become much severe. 

Now it is very difficult to arrange for any realistic theoretical set-up where the $f$ is close to $M_{Pl}$ \cite{Banks:2003sx}\footnote{A pNGB potential is used for natural inflation in \cite{Freese:1990rb}. The situation is similar in the context of inflation due to slow-roll conditions, but the problem is severe as the data requires $f > 3.5 M_P$ \cite{Ade:2013uln}.}. In our work, we consider this theoretical obstacle seriously and look for its possible resolutions. The $f$ parameter essentially controls the slope of the potential and larger $f$ makes the potential flatter which is preferred by the data.  A scalar field with a canonical kinetic energy term rolls down the potential too fast to be observationally consistent unless we let the field roll from the top of the potential.  %Borrowing from the original idea of N-flation \cite{Dimopoulos:2005ac}, 
It was suggested that $N$ pNGB fields can also collectively drive the present epoch of cosmic acceleration, where $f$ for individual pNGB can be much lower than the $M_P$, but the effective $f_{eff} = \sqrt{N} f$ can be easily close to $M_P$ making it observationally consistent \cite{Kaloper:2005aj}. 

In this work, we propose two well motivated modification to the single field pNGB potential, and  show that even smaller values of the $f$ being much smaller than $M_{Pl}$, the parameter is observationally consistent for any reasonable choice of initial condition for the field values. The central idea for both these modifications is to add extra terms (motivated by particle physics considerations) in the action that effectively increases the friction in the expanding Universe governed by Friedman equation. 

Among several modifications to the Einstein's gravity, a particularly well motivated class of models is the Galileon gravity based on the symmetry of the Galileon field $\pi$ as $\pi \rightarrow \pi + a + b_{\mu}x^{\mu}$ where $a$ and $b_{\mu}$ are constants \cite{Nicolis:2008in, Deffayet:2009wt}\footnote{Galileon theories are subclass of the most general theories proposed by Horndeski \cite{Horndeski:1974wa} with nonminimal interactions between the scalar field and gravity that give second order field equations in four dimensions.}.  This symmetry is motivated by the effective field theory of the decoupling limit of the DGP model \cite{Dvali:2000hr}. But in contrast to the DGP model, the Galileon theories are free of ghost as its equations of motions have derivatives up to second order. Based on this symmetry, five different field Lagrangians can be written based on the orders of the field derivatives. In this paper, in addition to the standard kinetic energy term of the pNGB field, we also consider the term that involves fourth derivatives of the field, namely $\mathcal{L}_3 = (\Box \phi)(\nabla \phi)^2/M^3$. As we will show, addition of this term modifies the Hubble equation in such a way that it increases the Hubble friction for the field $\phi$.

When the idea of Galieleon invariance is generalised to curved background, an additional term also can be added in the standard Einstein-Hilbert action: $\mathcal{L} = -\frac{1}{2M^2}G^{\mu\nu}
\partial_\mu \phi \partial_\nu \phi$, where $G^{\mu \nu}$ is the Einstein tensor \cite {Germani:2010hd}, \cite{Germani:2011bc}. Again, we will see that this term also effectively increases the Hubble friction. In the present paper we will discuss the effects of these two terms in pNGB dark energy models and confront those with the latest SN, CMB and BAO data.

This paper is organised as follows. In the next section we review 
the standard PNGB dark energy. and latest constraints on its parameter
space. In section III, we will outline the procedures of confronting dark energy models with recent SN, CMB and BAO data, and will confront the standard pNGB dark energy with canonical kinetic energy with data. This needs to be compared with the earlier work in \cite{Dutta:2006cf}. In section IV, we will outline the ideas in resolving the ``high-$f"$ problem and confront those with data. It would be clear that even small values of the $f$ parameter is well suited with observations.  In the last section we will conclude. 

\section{pNGB dark energy}
\label{review}
Once it is assumed that the source of dark energy is the existence 
of a quintessence scalar field, there are effectively infinite number 
of potentials that can serve the job required by observations \cite{Sahni:2004ai}. 
Moreover, 
many of these phenomenological potentials come with several parameters 
(often non-compact), leaving us with ample opportunities to fit the data. 
But a very important desirable property of the quintessence potential is 
that the potential must be stable under quantum corrections originating 
from the couplings to other quantum fields. To say it other way, it is not easy to write 
down a potential for a scalar field whose mass can be kept at $10^{-33} eV$. 
The rescue can 
come from a symmetry, in this case, it is a shift symmetry: 
$\phi \rightarrow \phi + c$. Exact shift symmetry allows only the constant 
potential with zero mass. But the symmetry can be broken in controlled 
way to give the field a suitably light mass. 

The above mentioned idea can be realised in the particle physics when a global $U(1)$ symmetry is spontaneously broken, giving rise to a massless Goldstone boson. 
The spontaneous breaking of a global symmetry gives rise to two modes namely
the radial modes that get massive and the angular modes which 
remains massless at the spontaneous symmetry breaking energy scale. These massless angular modes are called the NGBs.
%After sponteneous breaking of the global symmmetry, 
Now these spin-0 massless NGBs acquire masses (making those pseudo-NGBs) when there is another soft 
explicit breaking of the global symmetry at a lower energy
scale compared to the spontaneous symmetry breaking scale. The pNGB potential is  
characterised by these two symmetry 
breaking scales, spontaneous global symmetry breaking scale $f$
and explicit global symmetry breaking scale $\mu$, and it is given by 
\begin{eqnarray}
V(\phi)&=& \mu^4\left[1+\cos\left(\frac{\phi}{f}\right)\right]\,\,.
\label{pngb-pot}
\end{eqnarray}
The value of $f$ determines the steepness of the pNGB potential. As the value of $f$ increases, the pNGB potential becomes flatter. In fact $f \rightarrow \infty$ corresponds to the exact shift symmetric constant potential. 
pNGBs were first proposed  in the context of natural inflation \cite{Freese:1990rb}, and then it was subsequently extended for the case of dark energy \cite{Frieman:1995pm}.

%Slow-rolling scalar field models serve as good candidates of dark energy
%and also the alternatives to the cosmological constant. The flatness
%of the potential with respect to the field is one of the features
%which a scalar field potential must have to give rise to the slow-rolling
%situation which in turn produces the late time accelerated expansion 
%of the universe.
%The scalar field potential also have to be stable under radiative corrections.
%Another feature that a scalar field must have, is the
%hidden nature of its in local physics i.e., the scalar field must hide
%itself in the local gravity tests which requires the chamelion 
%mechanism \cite{Jain:2010ka} to come into play.
%These necessary conditions for a scalar field to act as dark energy
%requires huge fine tunning of the parameter space of the scalar
%field models. Also there are very few scalar field potential
%which arises from fundamental theory or arises in the particle physics
%sector of the standard particle physics model.
%In this regard, it was the PNGB potential which was proposed as dark 
%energy which has all these necessary features to give rise to 
%a viable dark energy model.
We consider the dynamics of a pNGB quintessence field with potential $V(\phi)$ of Eq. \eqref{pngb-pot} in a flat Friedmann-Robertson-Walker Universe governed by the Einstein's equations. The Lagrangian for a canonical pNGB field in this case looks like
\begin{eqnarray}
\cal{L} &=& \frac{M_{pl}^2}{2} R - X -V(\phi)
\end{eqnarray}
where $X=\frac{1}{2}\partial_\mu\phi\partial^\mu\phi$. The equations of motion in the late universe
containing matter are given by the Friedmann equation 
\begin{eqnarray}
H^2&=& \frac{1}{3 M_P^2}\left(\frac{1}{2}\dot\phi^2+V(\phi) +\rho_m\right)
\,\,,\label{fried1}
%2\dot H+3 H^2 &=& -8\pi G\left(\frac{1}{2}\dot\phi^2-V(\phi)\right)\,\,.
%\label{fried2}
\end{eqnarray}
and Klein Gordon equation for the scalar field
\begin{eqnarray}
\ddot\phi+3H\dot\phi+\frac{dV}{d\phi}&=&0\,\,,
\label{pngb-eos}
\end{eqnarray}
where $H = \dot a/a$ is the Hubble constant and $\rho_m$ is the energy density of the non relativistic matter. Note that the effects of radiation energy density is negligible for the late time cosmology that we are concerned with. We use these equations to obtain the evolution of the scalar
field and Hubble parameter to constrain the model parameters. We will compare the latest observational
constraints with the earlier work done in \cite{Dutta:2006cf}. 

\section{$\chi^2$ analysis of Observational data}
\label{Obs}
We use the latest type Ia supernovae data, Cosmic microwave background
shift parameter data and baryon acoustic oscillation data to constrain
the two different models discussed above.

%\subsection{SNe Ia data from Union 2.1}

Type Ia supernovae are considered to be the standard candles in astrophysics.
Measurements of luminosity distance ($d_L$) of 
the type Ia supernovae
with their redshifts happened to be the first probe \cite{Perlmutter:1998np}
of the discovery 
that
the Universe is undergoing an accelerated phase of expansion in the
present epoch.
The lateset compilation of Union 2.1 SNe Ia data has been performed 
by Suzuki \etal \cite{Suzuki:2011hu} and we use this set of 580 data points to
constrain our model parameters space.
The distance modulus is defined as
\begin{eqnarray}
\mu(z)&=& 5 \log_{10}(D_L(z))+\mu_0,
\end{eqnarray}
where $D_L(z)=H_0 d_L(z)/c$ ($c$ is the speed of light in vacuum) and
$\mu_0=42.38- 5 \log_{10} h$ with $H_0=100h ~{\rm Km Sec^{-1} Mpc^{-1}}$.
$\chi^2_{\rm SNe}$ is defined as
\begin{eqnarray}
\chi^2_{\rm SNe}(p_s)&=& \sum_i 
\left[\frac{\mu_{\rm obs}(z_i)-\mu_{\rm theo}(z_i,p_s)}{\sigma_i}\right]^2\,\,,
\end{eqnarray} 
where $p_s$ correspond to the model parameters that are constrained with
the type Ia supernovae data. We marginalise the $\chi^2_{\rm SNe}$ over the
nuisance parameter $\mu_0$ and use that marginalised  $\chi^2_{\rm SNe}$
for the data analysis perpose.
%
%Marginalising over the nuisance parameter $\mu_0$, we obtain the marginalised
%$\chi^2_{\rm SNe}$ as
%\begin{eqnarray}
%\chi^2_{\rm SNe}&=& A-\frac{B^2}{C}\,\,,
%\end{eqnarray}
%where $A,~B$ and $C$ are given by
%\begin{eqnarray}
%A &=&\sum_i 
%\left[\frac{\mu_{\rm obs}(z_i)-\mu_{\rm theo}(z_i,p_s,\mu_0=0)}{\sigma_i}
%\right]^2\,\,,\\
%B &=&\sum_i 
%\left[\frac{\mu_{\rm obs}(z_i)-\mu_{\rm theo}(z_i,p_s,\mu_0=0)}{\sigma_i}
%\right]\,\,,\\
%C &=&\sum_i \frac{1}{\sigma_i^2}\,\,.
%\end{eqnarray}

%\subsection{Cosmic microwave background shift parameter data}

Cosmic microwave background shift parameter $R$ is extracted from the
first peak in the cosmic microwave background temparature anisotropy
plot. This is more or less is a model independent parameter that is 
greatly used in constraining the dark energy models. Shift parameter is 
defined as 
\begin{eqnarray}
R(z_*)&=& (\Omega_m^0 H_0^2)^{1/2}\int_0^{z_*}\frac{dz}{H(z)}\,\,,
\end{eqnarray}
where $z_*$ is the radiation-matter decoupling redshift.% which is calculated as
%\begin{eqnarray}
%z_* &=& 1048[1+0.00124(\Omega_b^0h^2)]^{-0.738}
%\left[1+g_1(\Omega_m^0h^2)\right]^{g_2}
%\,\,,
%\end{eqnarray}
%where $g_1$ and $g_2$ are given by
%\begin{eqnarray}
%g_1 &=& 0.0783(\Omega_b^0h^2)^{-0.238}\left(1+39.5(\Omega_b^0h^2)^{-0.763}
%\right)^{-1}\,\,,\\
%g_2 &=& 0.560\left(1+21.1(\Omega_b^0h^2)^{1.81}\right)^{-1}\,\,.
%\end{eqnarray}
%
$\chi^2_{\rm CMB}$ is defined as
\begin{eqnarray}
\chi^2_{\rm CMB}&=&\left[\frac{R(z_*,p_s)-R}{\sigma_R}\right]^2\,\,.
\end{eqnarray}
We have used the CMB shift parameter %both from WMAP 9 year results
%$R=1.7$
%as well as 
from the Planck results $R=1.7499\pm 0.0088$ at the
decoupling redshift $z_*=1090.41$ \cite{Planck:2013nga}.

%\subsection{Baryon accoustic oscillation data}

At the very high energy of the early Universe, 
the baryons are simultaneously acted upon by the two oppositely
directed forces namely attractive gravitational 
force which tries to take them closer and the radiation pressure
which tries to take them away from each other. As a result of 
these two opposite
forces an oscillation comes into play 
in the baryon photon plasma and the disturbence
travel through the baryon photon medium with a sound speed which is close to
the speed of light. As the Universe cools down the sound speed drops down
and after the radiation matter decoupling the sound speed drops down to
zero and
the disturbance gets frozen in the large scale structure. 
This phenomenon is known as the baryon acoustic oscillation (BAO) and  
is observed as the excess number of galaxies at a certain length scale. 
This 
is been being measured by the Sloan Digital Sky Survey (SDSS)
in the form of two point galaxy correlation function \cite{Tegmark:2006az}.
We use the BAO data of $\frac{d_A(z_\star)}{D_V(Z_{BAO})}$ 
\cite{Percival:2009xn,Beutler:2011hx,Jarosik:2010iu,Blake:2011en}, 
where $z_\star$ is the decoupling redshift given by $z_\star \approx 1091$, 
$d_A$ is the comoving angular-diameter distance given by 
$d_A(z)=\int_0^z \frac{dz'}{H(z')}$ 
and $D_V(z)=\left(d_A(z)^2\frac{z}{H(z)}\right)^{\frac{1}{3}}$. 
%Data required for
%the analysis is depicted in table \ref{baodata}. 
We calculate $\chi_{BAO}^2$ as described in Ref. \cite{Giostri:2012ek}, 
where it is defined as,
\begin{equation}
 \chi_{BAO}^2=X_{BAO}^T C_{BAO}^{-1} X_{BAO}\,\,,
\end{equation}
where
\begin{equation}
X_{BAO}=\left( \begin{array}{c}
        \frac{d_A(z_\star)}{D_V(0.106)} - 30.95 \\
        \frac{d_A(z_\star)}{D_V(0.2)} - 17.55 \\
        \frac{d_A(z_\star)}{D_V(0.35)} - 10.11 \\
        \frac{d_A(z_\star)}{D_V(0.44)} - 8.44 \\
        \frac{d_A(z_\star)}{D_V(0.6)} - 6.69 \\
        \frac{d_A(z_\star)}{D_V(0.73)} - 5.45
        \end{array} \right)
\end{equation}
and the inverse covariance matrix $C_{\rm BAO}^{-1}$ is given in the Ref.
\cite{Giostri:2012ek}.

%\begin{widetext}
%\begin{align}
%C^{-1}=\left(
%\begin{array}{cccccc}
% 0.48435 & -0.101383 & -0.164945 & -0.0305703 & -0.097874 & -0.106738 \\
% -0.101383 & 3.2882 & -2.45497 & -0.0787898 & -0.252254 & -0.2751 \\
% -0.164945 & -2.45499 & 9.55916 & -0.128187 & -0.410404 & -0.447574 \\
% -0.0305703 & -0.0787898 & -0.128187 & 2.78728 & -2.75632 & 1.16437 \\
% -0.097874 & -0.252254 & -0.410404 & -2.75632 & 14.9245 & -7.32441 \\
% -0.106738 & -0.2751 & -0.447574 & 1.16437 & -7.32441 & 14.5022
%\end{array}
%\right).
%\end{align}
%\end{widetext}

%\begin{center}
%\begin{table*}
%\begin{tabular}{|c||c|c|c|c|c|c|}
%\hline
% $z_{BAO}$  & 0.106  & 0.2 & 0.35 & 0.44 & 0.6 & 0.73\\
%\hline \hline
% $\frac{d_A(z_\star)}{D_V(Z_{BAO})}$ &  $30.95 \pm 1.46$ & $17.55 \pm 0.60$ & $10.11 \pm 0.37$ & $8.44 \pm 0.67$ & $6.69 \pm 0.33$ & $5.45 \pm 0.31$  \\
%\hline
%\end{tabular}
%\caption{Values of $\frac{d_A(z_\star)}{D_V(Z_{BAO})}$ for different values of $z_{BAO}$.}
%\label{baodata}
%\end{table*}
%\end{center}

%\subsection{Combined analyses of the data sets}

We perform a combined analysis of all the data sets together by making a
combined $\chi^2_{tot}$ given by,
\begin{eqnarray}
\chi^2_{tot}&=&\chi^2_{SN}+\chi^2_{CMB}+\chi^2_{BAO}\,\,.
\end{eqnarray}
We minimize this $\chi^2_{tot}$ and find best fit values as well
as the $2\sigma$ C.L.s for the model parameters.

\section{Constraining standard pNGB dark energy}
\label{pNGB}
Using the above mentioned analysis techniques in confronting dark energy models with data, we will now reanalyse the observational viability of a standard pNGB dark energy field using the latest available data sets. In doing so, we solve Eq. \eqref{fried1} and Eq. \eqref{pngb-eos} numerically and calculate the Huuble parameter as a function of redshift. 

\begin{figure}[ht]
\centerline{\epsfxsize=3.4truein\epsfbox{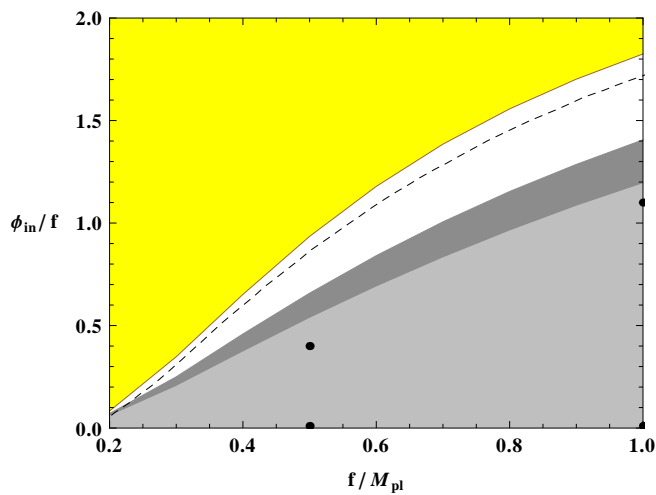}}
\caption{The upper left part marked with yellow is the region for which the Universe never evolves to $\Omega_{\phi}^{(0)} = 0.7$. The shaded areas at the bottom right part of the plot correspond to the $2\sigma$ and $5\sigma$ regions for $\Omega_{\phi}^{(0)} = 0.7$.  The dotted line marks $2\sigma$ confidence contour of the previous analysis \cite{Dutta:2006cf}. The black dots correspond to few representative points for which equation of state parameter $w(z)$ would be shown in Fig.~(\ref{fig4}).}
\label{fig1}
\end{figure}

%In this subsection we discuss the observational constraints on the standard
%pNGB dark energy. 
%To solve the model numerically i.e., 
%the Eqs. (\ref{fried1}) and (\ref{pngb-eos}), 
%we transform the quantities in dimensionless
%variables following the transformation rules 
%\begin{eqnarray}
%t\rightarrow H_0t\,\,,~~~~~
%\phi \rightarrow\frac{\phi}{M_{pl}}\,\,,~~~~~
%f\rightarrow \frac{f}{M_{pl}}\,\,.
%\label{trans-rules}
%\end{eqnarray}

A pNGB dark energy model has two parameters, namely $f$ and $\mu$ and two initial conditions $\phi_{in},~\dot\phi_{in}$. We use $\dot\phi_{in}=0$ in our calculations, and the assumption is reasonable considering that the large Hubble damping would typically make the field roll slowly at the initial stage. The present value of the dark energy density $\Omega_{\phi}^{(0)} \simeq 0.7$ can be traded with the value of $\mu$, and we are left with the parameter $f$ and the initial field value $\phi_{in}$. It is worth to remember that because of the periodic nature of the pNGB potential, $\phi_{in}$ can vary between $0$ and $2\pi f$, where the spontaneous symmetry breaking scale $f$ has a natural theoretical cut-off 
of the order of $M_P$. In principle, we can thus hope to exclude the whole model.
%and  and 
%the flatness condition replaces 
%one parameter among $f,~\mu,~\phi_{in}$ with the present
%epoch dark energy density $\Omega_\phi^0$. This essentially
%allows us to fit two parameters from the observational data 
%with the another parameter 
%held fixed.

%

\begin{figure}[ht]
\centerline{\epsfxsize=3.4truein\epsfbox{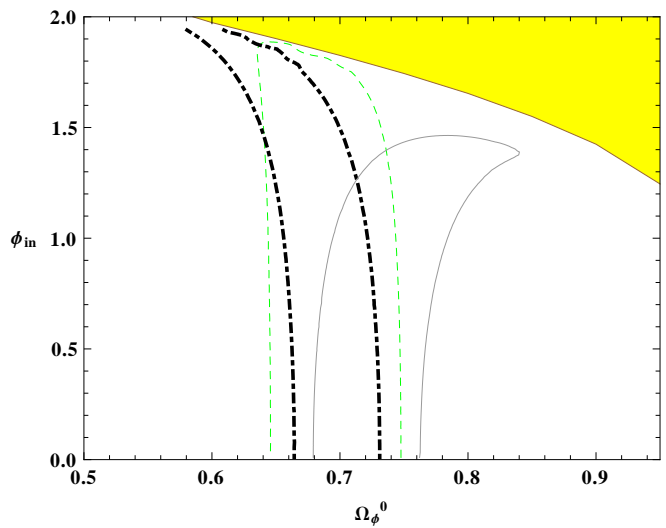}}
\caption{$2\sigma$ confidence contours in $\phi_{in}-\Omega_{\phi}^{(0)}$ plane for a fixed value of $f = M_{Pl}$. The yellow region is excluded by the fact that the corresponding $\Omega_{\phi}^{(0)}$ can not be reached. The smooth (black) line corresponds to the SN data, dotted-dashed (black) line is for CMB shift parameter and the dotted (green) line is due to the constraints coming from BAO data.}
\label{fig2}
\end{figure}

With $\Omega_\phi^{(0)}=0.7$ fixed, we plot the confidence contours in terms of $\phi_{in}~vs~f$ in Fig. (\ref{fig1}). The upper left shaded portion of the plot (yellow) is excluded by the fact that for those choice of parameters, the Universe can not evolve to a stage where $\Omega_\phi^{(0)}=0.7$. Looking at the $2 \sigma$ confidence contour marked by the light grey area, it is clear that $f$ close to $M_{Pl}$ is preferred observationally. Unless the initial field value is chosen carefully close to the top of the potential ($\phi_{in}\simeq 0$), $f \lesssim \mathcal{O}(0.5) M_{Pl} $ would be $2\sigma$ excluded. The nature of the confidence contours close to small $f$ and $\phi_{in}$ values tells that the statement remains more or less true with higher significance. In fact, the effective $\chi^2$ is minimum for $\phi_{in} \simeq 0$ for any values of $f$. But, it changes sharply when we increase $\phi_{in}$ value for smaller values of $f$. The field does not experience the steepness of the potential around $\phi_{in} \simeq 0$, and the effective dynamics is independent of $f$. This makes all values of $f$ equally probable for $\phi_{in} \simeq 0$.
As the field moves from $\phi_{in} \simeq 0$, it starts to experience
the slope of the potential and rolls down accordingly depending on $f$. Therefore if the potential
is steep enough the field will quickly evolve to the present epoch excluding the most part of the parameter space
in Fig. (\ref{fig1}). Note that the initial field values at the top of the potential is also very unlikely considering the quantum fluctuations produced by the earlier inflationary epoch at higher scale. Thus we are observationally pushed to a very high value of the spontaneous symmetry breaking scale $f \simeq M_{Pl}$, and it is difficult to accommodate from our theoretical understanding. This is what we call the ``high-$f$'' problem of pNGB quintessence \cite{Kaloper:2005aj}. In the next section, we will outline two possible resolutions of this problem and will discuss the observational implications.

We also compare our findings with the previous analysis derived in 
\cite{Dutta:2006cf}. We note that the $2\sigma$ region
has shrunk considerably where the dotted line marks the boundary of
$2\sigma$ contour of previous analysis. In fact, the present $5\sigma$ contour is within the previous $2\sigma$ contour. The comparison between these two figures indicates the 
improvement of the data points. The authors in the paper 
\cite{Dutta:2006cf} used 182 Gold SNe Ia data points 
\cite{Riess:2006fw} along with the 
CMB shift parameter constrain from WMAP 3 year data \cite{Wang:2006ts}, whereas
in this work we use 580 SNe data points from Union2.1 compilations
\cite{Suzuki:2011hu}, CMB shift parameter from PLANCK data \cite{Planck:2013nga}
and the baryon acoustic oscillation data from SDSS
\cite{Percival:2009xn,Beutler:2011hx,Jarosik:2010iu,Blake:2011en}. 
\begin{figure}[ht]
\centerline{\epsfxsize=3.4truein\epsfbox{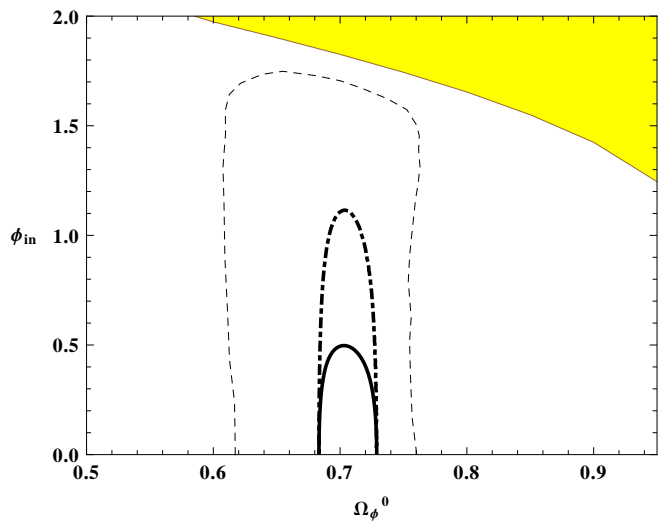}}
\caption{The dashed-dotted line (thick) corresponds to the $2\sigma$ confidence contour of joint analysis of SN, CMB and BAO data for $f = M_{Pl}$. This has to be compared with the previous analysis of \cite{Dutta:2006cf}, denoted by dotted curve (thin). The solid curve corresponds to the case when $f=0.5 ~M_{pl}$ (thick). 
%These points have been marked by black dots in Fig.~(\ref{fig1}).
}
\label{fig3}
\end{figure}

It is worth mentioning here that in the 
parameter space just bellow the theoretical bound, but outside the dotted line,
we have a small region for which $\Omega_\phi$ oscillates around 
$\Omega_\phi^{(0)}$. This corresponds to the case where the scalar
field crosses the potential minima and starts climbing up slowly. 
We can also identify the present epoch when the scalar
field is climbing up rather than rolling down. But it is clear from the plot that the parameters that allow this situation is
outside the  $5\sigma$ confidence limit from the present analysis\footnote{We thank Lorenzo Sorbo for making a clarifying point in this regard.}. 
%fig3beta0.eps   

In Fig.~(\ref{fig2}), we show the $2\sigma$ confidence regions in 
$\phi_{in}~vs~\Omega_\phi^{(0)}$ plane for $f=M_{pl}$. We plot the confidence contours for each individual observations, and the combined confidence contour (dot-dashed curve) is plotted in Fig.~\ref{fig3}. Even though all three different observations have vertical contours, but the combined plot produce a very stringent constraints in $\phi_{in}$ and $\Omega_\phi^{(0)}$ plane. This is more evident when the present constraints are compared with the old analysis of \cite{Dutta:2006cf} as shown in the dotted curve of Fig.~\ref{fig3}. The preferred value is around $\Omega_\phi^{(0)} \simeq 0.7$ with initial field value close to the top of the potential. There is no sign of evolution of pNGB field. When the similar plot is drawn for smaller values of $f$, the broad feature of the confidence contour (solid curve) remains the same, except that the contour shrinks along the $\phi_{in}$ axis, forcing the field to start closer to the top of the potential. This is in accordance with what we see in Fig. (\ref{fig1}).

\begin{figure}[ht]
\centerline{\epsfxsize=3.4truein\epsfbox{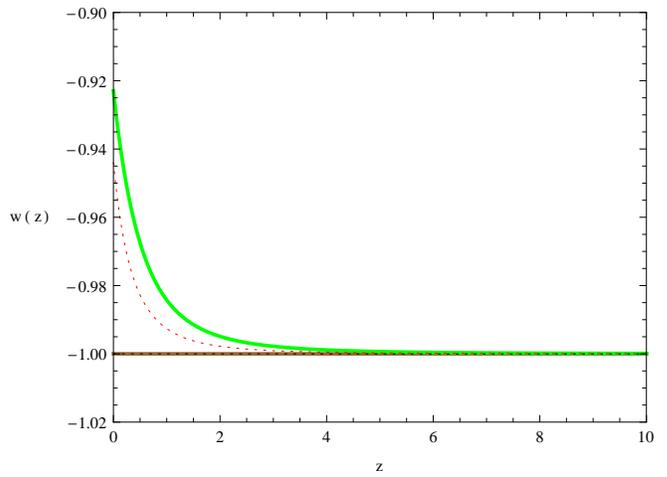}}
\caption{Variation of $w(z)$ with the redshift $z$ is shown here.
% \cite{Dutta:2006cf} and lateset bound on
%$\phi_{in}$ vs $\Omega_{\phi}^0$ plane.
The solid brown and solid green plots are for $\phi_{in}/f=0.01$
and $\phi_{in}/f=1.1$ with $f=M_{pl}$, $\Omega_{DE}=0.7$ respectively. 
 The dotted black and red plots are for the
$\phi_{in}/f=0.01$
and $\phi_{in}/f=0.4$ with $f=0.5~M_{pl}$, $\Omega_{DE}=0.7$ respectively.}
\label{fig4}
\end{figure}

\begin{figure}[ht]
\centerline{\epsfxsize=3.4truein\epsfbox{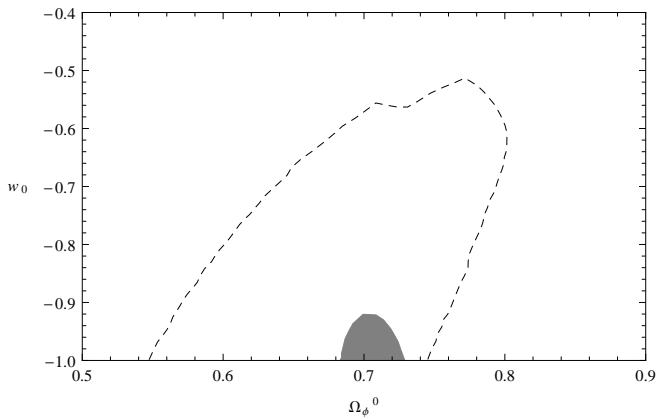}}
\caption{The plot shows the present value of the equation of state parameter as a function of present dark energy density. The shaded region shows the $2\sigma$ contours, 
whereas the dashed black curve is the previous $2\sigma$ constrain \cite{Dutta:2006cf}. }
\label{wz1}
\end{figure}

In Fig. (\ref{fig4}), we show the variation of the the dark energy 
equation of state $w$ as a function of the redshift $z$. 
%The solid green and brown plots are for $\phi_{in}/f=0.01$
%and $\phi_{in}/f=1.1$ with $f=M_{pl}$, $\Omega_{DE}=0.7$ respectively.
% The dotted black and red plots are for the 
%$\phi_{in}/f=0.01$
%and $\phi_{in}/f=0.4$ with $f=M_{pl}$, $\Omega_{DE}=0.7$ respectively.
The plots carry the clear signature that the best fit values mimick 
the cosmological constant behaviour. The deviations of $w$ from
$w=-1$ happens near the present epoch showing the thawing like behaviour of dark energy \cite{Linder:2007wa, Adak:2012bv}. Being stuck at the potential due to Hubble damping, the field shows very little evolution, and it is almost mimicking the cosmological constant. This is also clear from the plot in Fig. (\ref{wz1}) where the present value of the
dark energy equation of state parameter $w_0$ is shown against the 
present dark energy density. Even though this constraint was understood earlier, with recent data the parameter space has shrunk considerably. Within the $2\sigma$, the present equation of state parameter $w_0$ is constrained to be $w_0 < -0.92 $. 

The main finding in this section is that if a pNGB field is responsible for dark energy, the data favours large values of the $f$ parameter. Obviously the necessity of $f$ being large can be easily alleviated by choosing fine-tuned initial conditions. But we argue that this choice is unrealistic considering any prior inflationary epoch \cite{Kaloper:2005aj}. On the other hand, due to lack of theoretical understanding, it is difficult to construct models of pNGB where the decay constant $f$ is close to $M_{Pl}$. In the next sections, we will propose two well motivated particle physics modifications of the standard pNGB that can easily accommodate data even with smaller values of the $f$ parameter.  

\section{Helping hands}
\label{MP} 

In the previous section, we have seen that a pNGB quintessence field 
with canonical kinetic energy term is under strain with observational 
data. Only a fine-tuned initial conditions for the field with $f$ close 
to $M_{Pl}$ can mimic the observed expansion history of the Universe. 
On the other hand, having theoretically consistent model with large 
values $f$ is difficult to construct. The {\it helping hands} can 
come from the modifications to the Einstein's gravity. Here we propose 
two such modifications on the phenomenological ground 
motivated by the Galileon gravity and its 
generalisations. As we will see, both these examples will modify the 
Hubble equation enhancing the friction allowing small values $f$ 
observationally consistent. 
We note that a general Galileon field has usually non-derivative couplings 
to the matter sector. To evade the fifth-force constraints arising from 
these terms, the applicability of the Veinshtein mechanism is necessary 
\cite{Vainshtein:1972sx}. 
At the same time, the couplings to the matter those are 
of derivative types (if present) can circumvent the long-range force 
constraints by the usual Adler decoupling 
\cite{Carroll:1998zi} \cite{Kolda:1998wq} 

\subsection{Example 1}
\label{ex1}
We start with the following Lagrangian
\begin{eqnarray}
\cal{L} &=& \frac{M_{pl}^2}{2} R - X - F(\phi)X \Box \phi -V(\phi),
\end{eqnarray}
where the third term is just the $\mathcal{L}_3$ in the Galileon Lagrangian. Here
$\Box\phi=\frac{1}{\sqrt{-g}}
\partial_\mu\left(\sqrt{-g}\partial^\mu\phi\right)$. The Lagrangian is also a special case of the more general  ``Kinetic Gravity Braiding" dark energy models \cite{Deffayet:2010qz}
\begin{equation}
\mathcal{L} \supset K(\phi,X)+G(\phi,X)\Box \phi\,\,.
\end{equation}
These models though contains higher than the second derivative of the scalar fields, the equation of motion of the scalar 
fields do not contain more than second derivative of the scalar field and thus
these theories are free from Ostrogadsky ghosts \cite{ostro}. For simplicity we explore the situation where $F(\phi)$ is a constant
and is given by $F(\phi)=\frac{1}{M^3}$, where $M$ is a new mass scale.
 
\begin{figure}[h]
\centerline{\epsfxsize=3.4truein\epsfbox{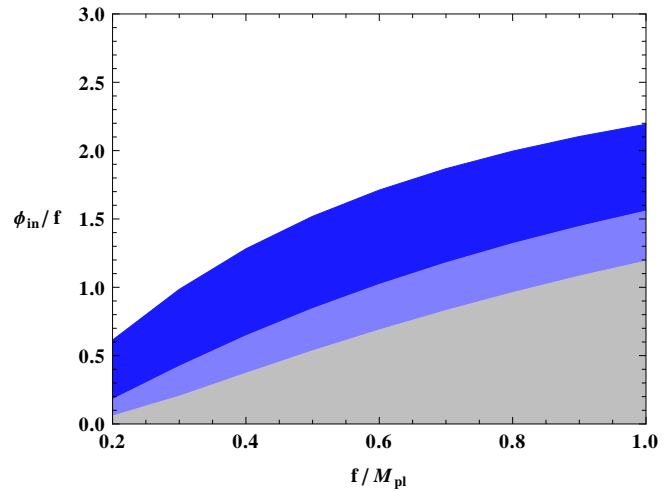}}
\caption{The allowed 2-$\sigma$ regions of 
$\phi_{in}/f$ vs $f/M_{pl}$ plane with $\Omega_\phi=0.7$ 
for different values of $\alpha$.
Light grey region is for $\alpha=0$ whereas darker blue and the darkest blue region
correspond to  $\alpha=1$ and  $\alpha=10$ respectively.}
\label{fig1beta-nonzero}
\end{figure}

The Friedmann equation and the equation of motion for the scalar field
are given by,
\begin{equation}
3 M_{pl}^2 H^2 = \frac{\dot{\phi}^2}{2}
\left(1+\frac{6}{M^3}H\dot\phi\right) +V(\phi) + \rho_m
\end{equation}
%\begin{equation}
%M_{pl}^2(2\dot H +3 H^2) = -\frac{\dot{\phi}^2}{2}
%\left(1-2 \frac{1}{M^3}\ddot\phi\right) +V(\phi)
%\end{equation}
\begin{equation}
\ddot \phi +3 H \dot\phi +3 \frac{1}{M^3} \dot\phi
\left(3 H^2 \dot\phi+\dot H \dot\phi +2 H \ddot\phi\right) +V'(\phi) = 0
\end{equation}
where $\rho_m$ is the matter density. To solve the model numerically, we transform the quantities in dimensionless
variables, and we have 5 parameters
$(\phi_{in},~\dot\phi_{in},~f,~\mu,~\alpha)$ where
$\alpha$ is a dimensionless parameter given by $\alpha=H_0^2 M_{pl}/M^3$, and parametrises the new physics. 
If the new physics scale $M=M_{pl}$, we find $\alpha \sim 10^{-124}$, without having any effect on the pNGB dynamics. 
The new mass scale should be such that the enough amount of
hubble friction is produced. For reasonable values of the dimensionless constant $\alpha \sim \mathcal{O}(1) - \mathcal{O}(100)$ the hubble friction makes the field roll down along the pNGB potential slowly independent of the steepness of the potential.

\begin{figure}[h]
\centerline{\epsfxsize=3.4truein\epsfbox{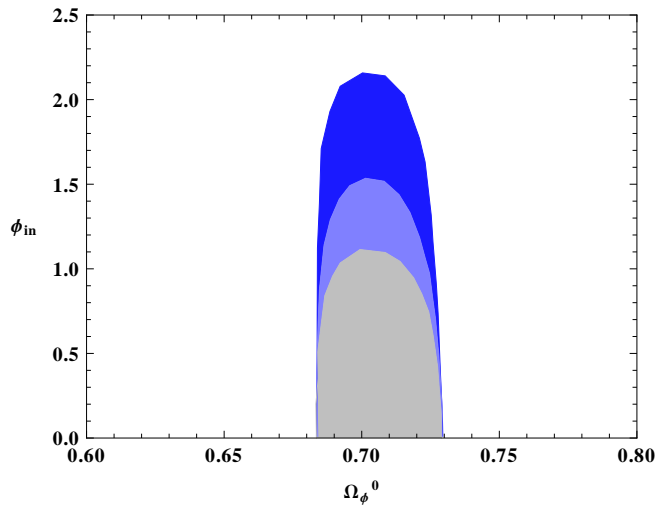}}
\caption{Plot showing the allowed 2-$\sigma$ regions of
$\phi_{in}$ vs $\Omega_{\phi}^0$ plane  with $f=M_{pl}$
for different values of $\alpha$.
Light grey region is for $\alpha=0$ whereas darker blue and the darkest blue region
correspond to  $\alpha=1$ and  $\alpha=10$ respectively.}
\label{fig3beta-nonzero}
\end{figure}

Assuming $\dot\phi_{in}=0$ and the flatness condition, we are left with only four
parameters $(\phi_{in},~f,~\Omega_\phi^0,~\alpha)$ where $\alpha$ parameterize the increased Hubble friction. $\alpha  = 0$ corresponds to the standard pNGB case as discussed in section \ref{pNGB}.
We choose different values of $\alpha$ with $\Omega_{\phi}^{(0)} = 0.7$ and plot the 2-$\sigma$ contours in the $\phi_{in}-f$ plane in Fig. \ref{fig1beta-nonzero}. 
%and use the other three
%parameters to put the observational constrains on the model.
%In 
%we show the $2\sigma$ contours for different values of $\alpha$.
%Standard pNGB model is given by $\alpha=0$. Other two values
%corresponds to $M=(H_0^2M_{pl})^{1/3}$ and $M=\frac{1}{10}(H_0^2M_{pl})^{1/3}$
%for $\alpha=1$ and $\alpha=1000$ respectively. This gives us
%some idea about the new mass scale $M$ which makes the KGB term 
%works in the theory to resolve the ``high $f$ problem" of pNGB
%potential.
Increasing the value of $\alpha$, the allowed region expands and smaller values of $f$ become equally favoured by the observations. At the same time, the field need not necessarily start to roll from the top of the potential. This implies that 
those initial values of $\phi_{in}$ which are disfavoured by the observations
in standard pNGB case ($\alpha=0$), becomes allowed as the values of $\alpha$  
increases. This is also clear from the Fig. \ref{fig3beta-nonzero} which has been plotted for $f = M_{Pl}$.

\subsection{Example 2}
\label{ex2}
In the following, we show another example that can do the similar job as like the previous example. The Lagrangian in this case is given by 
\begin{eqnarray}
\cal{L} &=& \frac{M_{pl}^2}{2} R -\frac{1}{2}g^{\mu\nu}
\partial_\mu \phi \partial_\nu \phi 
+\frac{1}{2M^2}G^{\mu\nu}
\partial_\mu \phi \partial_\nu \phi -V(\phi)\nonumber\\
\end{eqnarray}
where $G^{\mu \nu}$ is the Einstein tensor, and $M$ is a mass scale not necessarily linked to the previous example. The above modifications to the Einstein gravity was proposed in \cite{Germani:2010hd} in the context of inflation. In the context of modified gravity theories, the effects of the term have been studied in \cite{Adak:2013vwa}, but with different potentials and for different motivations. The phenomenology of the Lagrangian in the context of inflation was discussed in \cite{Maity:2012dx}.
Friedmann equations and the equation of motion for the scalar field
are given by,
\begin{eqnarray}
3M_{pl}^2 H^2&=&\frac{1}{2}\dot{\phi}^2+V(\phi)+\frac{9}{2M^2}H^2\dot\phi^2
+\rho_m,\\ 
M_{pl}^2(2\dot H +3 H^2) &=&-\frac{1}{2M^2}\dot\phi^2+V(\phi)
\nonumber\\&&
+\frac{1}{2M^2}\left((3H^2+2\dot H)\dot\phi^2+4H\dot\phi\ddot\phi\right)\,\,,
\end{eqnarray}
\begin{eqnarray}
\ddot \phi +3 H \dot\phi + \frac{1}{2M^2}\left(6H^2\ddot\phi+18H^3\dot\phi
+12H\dot H\dot\phi\right) +V'(\phi) &=& 0\,\,.\nonumber\\
\end{eqnarray}
where $\rho_m$ is the matter density. It is important to note that there are no higher derivative terms in the equations of motions. 
\begin{figure}[h]
\centerline{\epsfxsize=3.4truein\epsfbox{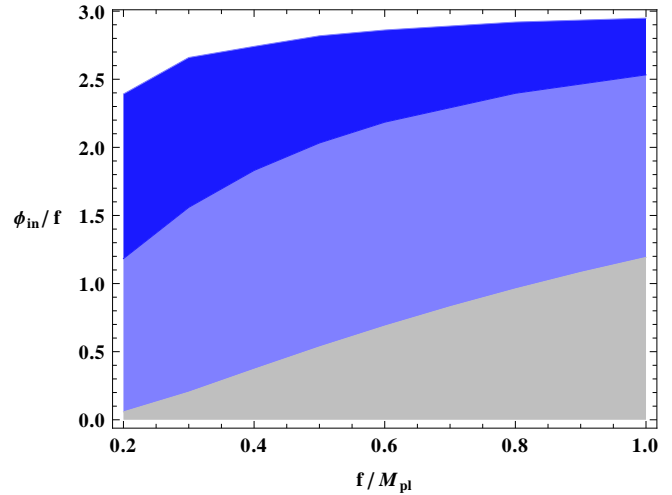}}
\caption{Plot showing the allowed regions of
$\phi_{in}/f$ vs $f/M_{pl}$ plane  with $\Omega_X^0=0.74$
in the $2\sigma$
C.L. for different values of $\beta$.
Light grey region is for $\beta=0$ whereas darker blue and the darkest blue region
correspond to  $\beta=1$ and  $\beta=10$ respectively.}
\label{fig1alpha-nonzero}
\end{figure}

Like in the previous example, here also we use a dimensionless parameter $\beta=H_0^2/M^2$ that captures the effect of the non-minimal coupling of the Eienstein tensor to the kinetic term. The role of $\beta$ is similar to $\alpha$ in the previous example. The effect of non-zero $\beta$ has been shown in the Fig. \ref{fig1alpha-nonzero}. It is clear that for moderate values of $\beta$, a large portion of the parameter space becomes allowed. 
%In Fig. \ref{w01}, we show the equation of sate parameter $w$, and see that the extra friction term does not allow the field move much, mimicking almost a cosmological constant. 
%\begin{figure}[h]
%\centerline{\epsfxsize=3.4truein\epsfbox{w0.eps}}
%\caption{Plot showing the allowed regions of
%$w0$ vs $\Omega_{\phi}^0$ plane  with $f=M_{pl}$
%in the $2\sigma$
%C.L. The blue line is for standard pNGB and the red line is for $\alpha=1$ as discussed in section \ref{ex1}.}
%\label{w01}
%\end{figure}

\section{Discussions and Conclusions}

If observed cosmic acceleration is due to a dynamical degree of freedom, namely a quintessence field, a pNGB scalar field is the most suitable candidate. This is due to the the stability of its mass from quantum corrections, as well as its ability to evade the fifth force constraints. But once we confront a pNGB potential with cosmic data, we see that the data favours large value of the spontaneous symmetry breaking $f$-parameter, as well as initial field value at the top of the potential. On the other hand, it has been argued that constructing theoretically consistent model with $f \simeq M_{Pl}$ is very challenging. As we have seen in section IV, with  $f <0.1 M_{Pl}$, the field must start to roll exactly from the top of the potential. If we combine it with the physics prior to the dark energy, say inflation, these initial conditions are very unlikely due quantum fluctuations. Combining all these ingredients, it seems that a standard pNGB is under strain. Even though it was understood earlier \cite{Dutta:2006cf}, we see that the recent observational data has severely constrained the parameter space. In particular, the present analysis done in section IV includes the BAO data, and it has a significant effect in constraining the model. In summary, the observation pushes the model to have large values of $f$ being close to $M_{Pl}$ with initial field values close to the top of the potential. But both these requirements are difficult to meet if we take our theoretical prejudices seriously. 

The way out to the problem may come from the introduction of several pNGB fields driving quintessence. But we propose other alternatives with single pNGB driving quintessence. In particular, we work on two alternatives where the Einstein Hilbert action is modified. These modifications are motivated by the Galieleon cosmology and its generalisations. Even though the the Lagrangian involves higher derivative terms, the equations of motions are of second derivative. For both these examples the main goal is to increase the friction due to non canonical terms. We then confront these modified models with data to show that for $\mathcal{O}(1) - \mathcal{O}(10)$ values of the dimensionless constants, a large part of the parameter space become viable. 

From the values of $\alpha$ in Sec. \ref{ex1} and $\beta$ in Sec. \ref{ex2}, it is evident that the new mass scale $M$ has to be of the order of $H_0$ which is much smaller than the Planck scale $M_{Pl}$. In fact the modified pNGB reduces effectively to the standard pNGB when $M = M_{Pl}$. Unless we have a more fundamental theory, we can not know what might be the source of this mass scale, but it is clear that to have the significant effect in the cosmic expansion at the present epoch, it must be of the order of $H_{0}$. In our work, even though we had analysed two examples separately for simplicity, but the combined effect can also be analysed. All these modified Lagrangians are just subset of a larger class given by Hordenski, and dark energy phenomenology of these general Lagrangian have been analysed already. 

%Large K.E can also be ... effects of other terms of the Galileon gravity. 

%In this paper, we have shown that the a single pNGB field with 
%canonical kinetic energy term is under tension with data and 
%theoretical consistencies. We have proposed two modifications 
%to the Einstein gravity that alleviate the problem and confront 
%those modified models with data. 

In the standard pNGB dark energy case, a pNGB scalar field that gives rise 
to late-time cosmic acceleration,
couples to the matter derivatively \cite{Carroll:1998zi}. 
These interactions lead to a fifth-force that is suppressed by the 
Adler decoupling in the low momentum limit. On the other hand, the couplings 
of a Galileon scalar field to the matter is of non-derivative types 
$ S_m[\psi_m; e^{2\beta \phi/M_{pl}} g_{\mu\nu}]$ \cite{Ali:2012cv}, 
and to suppress the fifth-force in this case, it is required to invoke 
the Vainshtein mechanism. In our case, motivated by the Galileon theories, 
we have introduced 
$\frac{1}{M^3}X\Box \phi$ and 
$\frac{1}{2M^2}G^{\mu\nu}\nabla_\mu \phi \nabla_\nu \phi$
terms to resolve the ``high $f$'' issue of axionic decay constant. 
As these terms are motivated by the Galileon gravity, it is natural 
to expect that the couplings of the scalar field to the matter sector 
will contain both derivative or non-derivative interactions. The 
derivative couplings of the scalar field gives rise to fifth force 
which is highly suppressed by Adler
decoupling mechanism \cite{Carroll:1998zi} \cite{Kolda:1998wq}. 
The non-derivative couplings with matter 
(can arise from soft breaking of shift symmetry) can give rise 
to fifth-force constraints, but those are usually suppressed by 
the Vainshtein mechanism \cite{Ali:2012cv} \cite{Kase:2013uja}.

In this work we have studied the recent observational 
constraints on the pNGB dark energy. We find
that the low $f$ region ($f<M_{pl}$) is highly constrained 
and about to be ruled out by the observations.
In this situation, we offer two ``helping hands'' strictly on 
the phenomenological ground and study the effects of these terms 
in resolving the ``high $f$'' issue of the pNGB dark energy. 
These two terms are well motivated in the context of Galileon 
cosmology \cite{Ali:2012cv} as well as in the context of 
``Kinetic Gravity Braiding''
dark energy models \cite{Deffayet:2010qz} and have been studied earlier in the 
context of inflation \cite{Maity:2012dx} \cite{Germani:2010hd}. 
We have shown that in presence of these terms, 
low $f$ values go well with the observations.
This is the main result of this work in addition to revising the 
constraints with the latest available observational data. 
%Being a 
%pseudo-scalar, a pNGB field couples with matter derivatively and 
%follows the principle of Adler decoupling at low momentum
%\cite{Carroll:1998zi,Kolda:1998wq,Chiba:1999wt} 
%\cite{Brax:1999gp}. 
%Thus the fifth force constraints are weak, and it is not necessary 
%to invoke mechanisms like Vainstein \cite{Vainshtein:1972sx} 
%or Symmetron \cite{Hinterbichler:2010es} %Chameleon \cite{Khoury:2003aq}
%to evade fifth force
%constraints. Surely, the revised fifth force constraints 
%for a pNGB Galileon is
%an issue of complete study which is beyond the scope of this paper.
%that we do not attempt to do 
%in this work. 
%But we do use the effects of Galileon terms 
%(modifications) in the background dynamics to achieve small $f$ 
%being consistent with data. 
It would be really interesting to find 
a concrete theoretical construction where a pNGB can be a 
Galileon type field. We leave this effort for future work.

\section{Acknowledgement}
D.A is supported by the fellowship from SINP. K.D is partially supported by the Ramanujan Fellowship and Max Planck Society-DST Visiting Fellowship. We thank all the participants of the Friday Cosmology Discussions $@$ SINP for keeping us moving.

\bibliography{pNGB.bib}{}
\bibliographystyle{h-physrev3.bst}

%\begin{thebibliography}{99}
%\bibitem{sami}
%M. Sami
%\bibitem{Linder}
%E. V. Linder
%\bibitem{string-theory}
%M. Dine
%\bibitem{inflation}
%\bibitem{koushik}
%Koushik Dutta, Lorenzo Sorbo 
%\bibitem{sn}
%N. Suzuki
%\bibitem{wmap}
%E. Komatsu
%\bibitem{bao}
%D. Eisenstein
%\bibitem{debaprasad}
%Debaprasad Maity
%\bibitem{christiano-germany}
%Christiano Germany
%\bibitem{kaloper}
%Kaloper
%\bibitem{freeman}
%Freeman
%\bibitem{deffayet}
%C. Deffayet
%\bibitem{perlmutter}
%Sol perlmutter
%
%
%
%
%
%
%
%\end{thebibliography}
\end{document}